\documentclass[twocolumn,prb,showpacs,amsmath,amssymb,rotating]{revtex4}

\usepackage{graphicx}
\usepackage{dcolumn}
\usepackage{bm}
\usepackage{ulem}

\begin{document}

\preprint{}

\title{Weak Localization and Spin Splitting in Inversion Layers on p-type InAs}

\author{Christopher Schierholz}
\email{cschierh@physnet.uni-hamburg.de}
\author{Toru Matsuyama}%
\author{Ulrich Merkt}%
\author{Guido Meier}%
\affiliation{%
Institut f\"ur Angewandte Physik und Zentrum f\"ur
Mikrostrukturforschung, Universit\"at Hamburg, Jungiusstra\ss e
11, D-20355 Hamburg}%

\date{\today}
\begin{abstract}
We report on the magnetoconductivity of quasi two-dimensional
electron systems in inversion layers on p-type InAs single
crystals. In low magnetic fields pronounced features of weak
localization and antilocalization are observed. They are almost
perfectly described by the theory of Iordanskii, Lyanda-Geller and
Pikus. This allows us to determine the spin splitting and the
Rashba parameter of the ground electric subband as a function of
the electron density.
\end{abstract}


\maketitle

The multitude of new applications promised by use of the
electron's spin as a degree of freedom in addition to its charge
has led to growing interest in the area of spintronics
\cite{Wolf01}. For semiconductor based spintronics, the required
control of the electron spin is expected to be achieved via
spin-orbit interaction. In single crystals this interaction
originates from two terms: the bulk inversion asymmetry of the
crystal lattice \cite{Dresselhaus55} and the structure inversion
asymmetry \cite{Rashba84}. The latter term includes contributions
from the electric field and from the boundary conditions and is
commonly called Rashba term. For many heterostructures good
agreement between calculated Rashba parameters and experimentally
deduced values is reached \cite{Grundler00, Koga02}. In the case
of two-dimensional electron systems (2DES) in surface inversion
layers the spin-orbit interaction and the Rashba parameter are
still under debate \cite{Matsuyama00, Lamari01, Pfeffer03}.

Theoretical considerations \cite{Lommer88, AndradaESilva94} have
shown that the Rashba term is the main cause of the spin splitting
of electric subbands in InAs structures. This is supported by the
experiments of Luo \textit{et al.}~on InAs quantum wells
\cite{LuoBoth}. It has been demonstrated for various
he\-te\-ro\-struc\-tures \cite{Nitta97, Engels97, Grundler00} as
well as for inversion layers on p-type InAs single crystals
\cite{Matsuyama00} that the Rashba term can be influenced by an
external gate voltage. This was concluded from beating patterns in
Shubnikov-de Haas (SdH) oscillations at various gate voltages.
However, contrary to SdH experiments, all proposed spintronic
devices operate at zero or low magnetic fields. Calculations by
Lommer \textit{et al.}~showed that not only the magnitude but even
the sign of the spin splitting can change between low and high
magnetic fields for AlGaAs/GaAs heterostructures \cite{Lommer88}.
For spintronic applications it is thus necessary to determine the
spin splitting at near-zero magnetic fields.

Dresselhaus \textit{et al.}~\cite{Dresselhaus92} showed that the
destruction of weak antilocalization by small magnetic fields can
be used to determine the strength of the spin-orbit interaction in
GaAs. These authors analyzed their data with the theory of Hikami,
Larkin, and Nagaoka \cite{Hikami80}. Kawaguchi \textit{et
al.}~\cite{Kawaguchi87} found that this theory could not correctly
reproduce the weak antilocalization observed in inversion layers
on p-type InAs. Koga \textit{et al.}~\cite{Koga02} showed that a
quantitative description of weak antilocalization in
In\-Al\-As/In\-Ga\-As/In\-Al\-As heterostructures is possible with
the model developed by Iordanskii, Lyanda-Geller, and Pikus (ILP)
\cite{Iordanskii94, Knap96}. An\-a\-ly\-ses of weak
antilocalization using ILP theory have been performed on
AlGaAs/AlInAs quantum wells \cite{Knap96}, Al\-Ga\-As/Ga\-As
p-type quantum wells \cite{Pedersen99} and
Al\-In\-As/Ga\-In\-As/Al\-In\-As quantum wells \cite{Koga02}.
However, this approach so far has not been used to analyze the
magnetoconductance of 2DES on bulk narrow-gap semiconductors.

\begin{figure}[t]
\vspace{4mm}
\includegraphics[width=8.2cm]{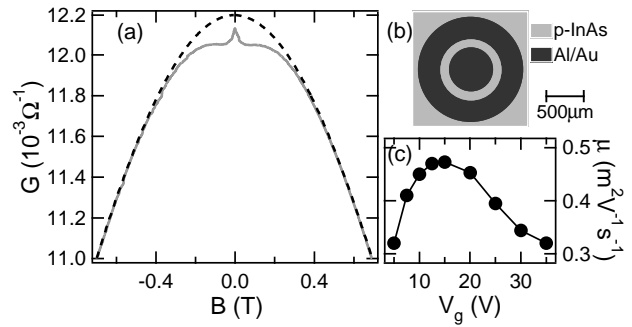}
\vspace{-2mm} \caption{(a) Experimental magnetoconductance for a
gate voltage of $\rm{V_{g}= 15\ V}$ at T=2.65 K (solid line) as
well as a calculated Drude parabola (dashed line). (b) Schematic
of the sample geometry. The radius of the inner electrode is
$\rm{300\ \mu m}$, the channel length $\rm{100\ \mu m}$, and the
radius of the outer electrode $\rm{700\ \mu m}$. (c) Electron
mobility versus gate voltage.} \label{Fig1}
\end{figure}

We have performed measurements on Zn doped p-type InAs (100)
single crystals with an acceptor concentration of approximately
$\rm{N_{A}=2\cdot10^{17}\ cm^{-3}}$. Band bending leads to a 2DES
confined in the approximately triangular asymmetric potential well
characteristic of a natural surface inversion layer \cite{Ando82}.
The narrow band gap of InAs results in strong spin-orbit
interaction, with the Rashba term that originates from the
asymmetry of the potential well dominating the spin splitting of
the electric subbands \cite{Lommer88}. Our samples are
field-effect transistors in Corbino geometry. The electrodes are
defined by optical lithography and deposited by thermal
evaporation. They consist of 35 nm thick Aluminium passivated by
10 nm of Gold. The entire structure is covered by a 340 nm thick
$\rm{SiO_2}$ insulator onto which a gate covering the
semiconductor channel is patterned. All measurements are performed
in lock-in technique at a temperature of 2.65 K in a $\rm{^{4}He}$
cryostat.

The measured magnetoconductances show strong parabolic signatures
as can be seen for fields above $\rm{300\ mT}$ in
Fig.~\ref{Fig1}(a). This classical Drude magnetoconductivity
$\rm{G(B)=G_{0}\cdot(1+\mu^{2}B^{2})^{-1}}$ stems from the Corbino
geometry depicted in Fig.~\ref{Fig1}(b). A parabola fit to the
experimental data yields the zero-field conductivity $\rm{G_0}$ as
well as the electron mobility $\rm{\mu}$. The mobility as a
function of the applied gate voltage is shown in
Fig.~\ref{Fig1}(c). It is characteristic for a
metal-oxide-semiconductor (MOS) transistor: surface scattering at
low gate voltages and ionized impurity scattering at high gate
voltages lead to reduced mobilities \cite{Ando82}. The inversion
threshold of the sample lies at a gate voltage of $\rm{V_{g}}\cong
1.8~V$. The total electron density $\rm{n_s}$ is determined from
SdH oscillations of the source-drain resistance. We find a linear
dependence $\rm{n_{s}[cm^{-2}]=1.46\cdot 10^{11} + 6.70\cdot
10^{10}~ V_{g}[V]}$ on the gate voltage.

\begin{figure}[t]
\vspace{4mm}
\includegraphics[width=8.2cm]{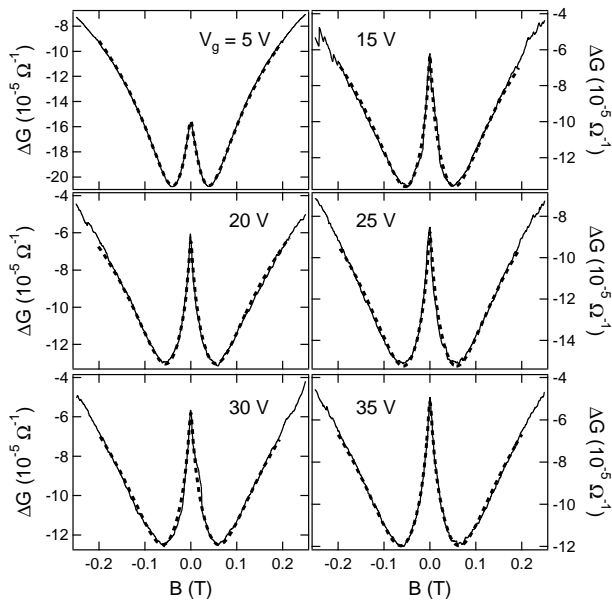}
\vspace{-2mm} \caption{Experimental magnetoconductances recorded
at 2.65~K for various gate voltages (solid lines). The classical
Drude conductance is subtracted. Theoretical conductances (dashed
lines) are calculated from the theory of Iordanskii,
Lyanda-Geller, and Pikus (Refs.~\onlinecite{Iordanskii94,
Knap96}).} \label{Fig2}
\end{figure}

The Drude parabola is subtracted from the magnetoconductance to
leave the quantum corrections originating from weak localization
and antilocalization remaining. Starting from zero-field, these
corrections at first exhibit a negative magnetoconductance
originating from spin-orbit interaction based weak
antilocalization followed by a positive magnetoconductance due to
weak localization. Figure~\ref{Fig2} shows results for various
gate voltages. The conductance corrections $\rm{\Delta G(B)}$ are
fit according to the ILP theory which provides an excellent
description. Considering the structure inversion asymmetry as
dominant origin of the zero-field spin splitting in our system and
thus neglecting the bulk inversion asymmetry ($\rm{\Omega_{3}=0}$
in Refs.~\onlinecite{Iordanskii94} and \onlinecite{Knap96}), only
the characteristic magnetic fields of inelastic and of spin-orbit
scattering
\begin{equation}\label{HphiHso}
   \rm{H_{\phi}={\hbar\over{4De\tau_{\phi}}} \quad and
   \quad H_{SO}={{\Delta_{0}^{2} \tau_{tr}}\over{8\hbar
De}}={1\over{8\hbar^{3} e \pi}}{({m^*}
\Delta_{0})^{2}\over{n_s}}\quad
   }
\end{equation}
are relevant in the fitting procedure \cite{Knap96}. In these
relations D is the diffusion coefficient, $\rm{\tau_{\phi}}$ and
$\rm{\tau_{tr}}$ are the inelastic and transport relaxation times,
respectively, and $\rm{\Delta_{0}}$ is the spin-splitting energy
in zero magnetic field.  The characteristic field $\rm{H_{tr}}$ of
transport scattering can be ignored in the fit as it only results
in a shift in $\rm{\Delta G(B)}$ \cite{Knap96,Koga02}. The
diffusion coefficient is given by $\rm{D={\tau_{tr}v_{F}^{2}\over
2}}$ with the Fermi velocity $\rm{v_{F}={\hbar k_{F}\over
{m^{*}}}}$, the Fermi wave vector $\rm{k_F}$, and the electron
effective mass $\rm{m^{*}}$.

We calculate the zero-field spin splitting $\rm{\Delta_{0}}$ from
Eq.~\ref{HphiHso} and the Rashba parameter from the relation
\cite{Koga02}
\begin{equation}\label{alphaWALHso}
   \rm{\alpha={\sqrt{\hbar^{3}eH_{SO}}\over{m^{*}}}}.
\end{equation}
Note that only the absolute value of the Rashba parameter can be
determined, not its sign. In case of inversion layers we do not
expect a change of sign as their asymmetric surface potentials do
not change their overall shape. The analysis is performed for the
effective mass $\rm{m^{*}}$ of the ground electric subband at the
Fermi energy. For high carrier densities $\rm n_{s}$ this mass is
determined from temperature dependent SdH oscillations. At the
inversion threshold the band edge mass of the ground subband of
$\rm{m_{0}^{*}=0.026 \cdot m_{e}}$ is used \cite{Oelting87}.
Hence, we find
\begin{equation}
\rm{m^{*}}=(0.026 + 0.012 \cdot 10^{-12}~n_{s}- 0.001 \cdot
10^{-24}~n_{s}^2)\cdot m_{e},\label{eq_eff_mass}
\end{equation}
with the electron density $\rm{n_{s}}$ in units of $\rm{cm^{-2}}$.

\begin{figure}[t]
\vspace{4mm}
\includegraphics[width=6.8cm]{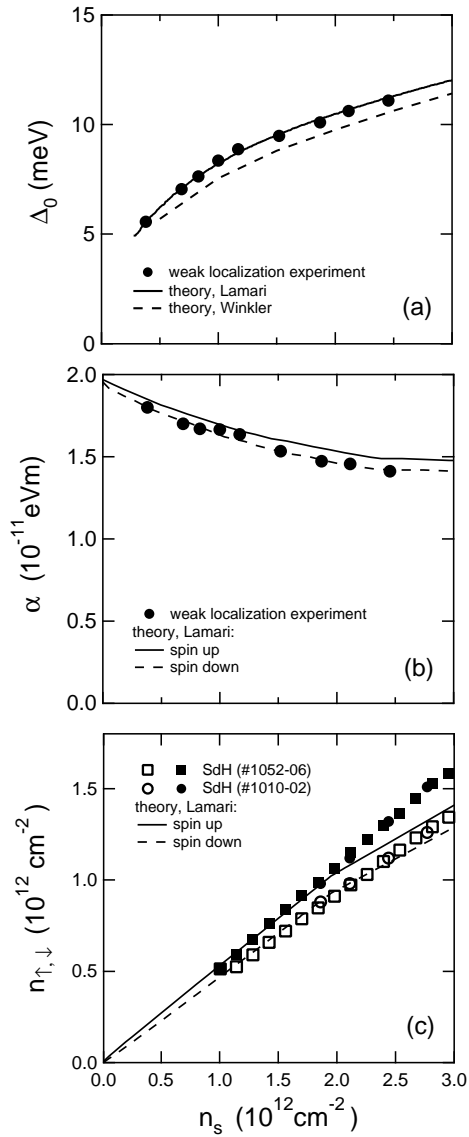}
\vspace{-2mm} \caption{(a) Spin-splitting energy at the Fermi
level. Theoretical data are taken from Lamari
(Ref.~\onlinecite{Lamari03}) and Winkler
(Ref.~\onlinecite{Winklerpc}). (b) Rashba parameter. Dashed and
solid lines are calculated for the spin-split subbands at the
Fermi level (Ref.~\onlinecite{Lamari03}). (c) Occupation of the
spin-split subbands. Open and closed symbols are data for
spin-down and spin-up states of the two different samples
indicated by circles and squares. Theoretical curves are taken
from Lamari (Ref.~\onlinecite{Lamari01}).} \label{Fig3}
\end{figure}

Figure \ref{Fig3}(a) shows the carrier-density dependence of the
spin-splitting energy $\rm \Delta_{0}$. The splitting increases
monotonically with the electron density $\rm{n_{s}}$ and is in
good agreement with band-structure
calculations~\cite{Lamari01,Lamari03,Winklerpc} displayed in the
same figure. The Rashba spin-orbit parameter $\rm{\alpha}$ is
shown in Figure \ref{Fig3}(b). The analysis of weak localization
experiments with the ILP theory yields one Rashba parameter for
each carrier density. Band-structure calculations yield different
Rashba parameters for the two spin subbands because of the slight
difference between the Fermi wave vectors for spin-up and
spin-down states. However, this difference is small and can be
neglected in the comparison with the weak localization data. A
decrease of the Rashba parameter with increasing carrier density
is observed. Again, the experimental results are in good agreement
with the values obtained from the calculations.
Figure~\ref{Fig3}(c) shows the occupations of the spin-split
subbands as determined from SdH oscillations
\cite{Schierholz02,Matsuyama00}. At carrier densities below
$2.0\cdot 10^{12}\ \rm{cm}^{-2}$ a good agreement between theory
and experiment is observed. The first excited subband is only
populated for carrier densities above about $2.3\cdot 10^{12}\
\rm{cm}^{-2~}$ \cite{KuerstenDiss}. In the theory for a doping
concentration of $1.8\cdot 10^{17}~\rm{cm}^{-3}$, population of
the first excited subband occurs for carrier densities above
$2.0\cdot 10^{12}\ \rm{cm}^{-2}$, leading to kinks in the plots.
This can be seen most easily in Fig.~\ref{Fig3}(c) for the spin-up
electrons.

Values of the Rashba parameter determined from SdH oscillations in
high magnetic fields \cite{Schierholz02,Matsuyama00} do not agree
with the multi-band calculations of Lamari \cite{Lamari01,
Lamari03} and Winkler \cite{Winklerpc} and the results from weak
localization reported here. While they are of the same order of
magnitude, they increase with increasing carrier density.

Using the effective mass from Eq.~\ref{eq_eff_mass} and the
subband carrier densities of Lamari displayed in
Fig.~\ref{Fig3}(c) the simple relation \cite{KuerstenDiss} $\alpha
= \hbar^2[\sqrt{4 \pi n_{\uparrow}} - \sqrt{4 \pi
n_{\downarrow}}]/m^*$ yields Rashba parameters that agree well
with those in Fig.~\ref{Fig3}(b). Hence the effective mass cannot
be responsible for the discrepancy between the results in low and
high magnetic fields. We currently do not understand why the
splitting of the spin-subband densities from SdH oscillations as
well as the increase thereof with growing carrier density are
larger than theoretically predicted.

The situation in heterostructures of low carrier density is
different, as a good agreement between theory and high-field
experiments is found \cite{Engels97}. This indicates that the
simple evaluation of beating patterns in SdH oscillations is not
sufficient for the high carrier densities in 2DES inversion layers
on p-type InAs crystals. In the carrier-density range well above
$10^{12}~\rm{cm}^{-2}$ the spin-split energies no longer exhibit
linear dispersion. Then band nonparabolicity and higher order
terms in the expansion of the Rashba parameter become decisive.
Hence it is important to consider the Rashba parameter as a
function of both the electric field as well as of the in-plane
wave vector \cite{AndradaESilva94,deAndrada97}. The wave vector
dependence exceeds the electric field contribution at high
electron densities and leads to a decrease of the Rashba
parameter. Obviously, for inversion layers on p-type InAs the
evaluation of the weak localization measurements provides more
reliable values for the Rashba parameter than that of SdH
oscillations.

To conclude, we have studied spin-orbit interaction in the surface
inversion layer on p-type InAs single crystals using weak
localization at near zero magnetic field. Excellent description of
the experimental data could be achieved through the theory of
Iordanskii, Lyanda-Geller and Pikus \cite{Iordanskii94, Knap96}.
We find the carrier-density dependence of the spin-splitting
energy $\Delta_{\rm 0}$ and the Rashba parameter $\alpha$ of the
ground electric subband to be in very good agreement with
multi-band calculations, but to differ from previous analyses in
which SdH beating patterns were evaluated.

\begin{acknowledgments}
We thank Saadi Lamari and Roland Winkler for valuable discussions
and for providing detailed theoretical data, Alexander Thieme for
software development. The authors gratefully acknowledge financial
support from the BMBF via the Verbundprojekt 13N8281 and from the
Deutsche Forschungsgemeinschaft via SFB 508.
\end{acknowledgments}

\bibliography{InAs-bulk_paper}

\newpage

\end{document}